\documentclass[11pt]{article}
\usepackage{amsfonts,latexsym,eucal,color,graphicx,epsfig,amssymb,amsmath,cite}
\newcommand{\be}{\begin{eqnarray}}
\newcommand{\ee}{\end{eqnarray}}

\newenvironment{minilinespace}{\baselineskip = 8mm}{}

\begin{document}

\begin{titlepage}

\begin{flushright}
{ \small
	arXiv:0803.3037 [hep-th]\\
	WU-AP/279/08
}
\end{flushright}

\vspace{1cm}

%---------------------------------------------------------------------%
\begin{minilinespace}
\begin{center}
	{
		\Large
		{\bf
			Liquid bridges and black strings in higher dimensions
		}
	}
\end{center}
\end{minilinespace}
\vspace{1cm}
%---------------------------------------------------------------------%

\begin{center}
Umpei Miyamoto$^{1,a}$ and Kei-ichi Maeda$^{2,b}$ \\
%%--------   Address  ------------------
\vspace{.5cm}
{\small \textit{
${}^1$
Racah Institute of Physics, Hebrew University, Givat Ram, Jerusalem 91904, Israel\\
${}^2$
Department of Physics, Waseda University, Okubo 3-4-1, Tokyo 169-8555, Japan
}}
\\
%%----------------------------------------
\vspace*{0.5cm}

%%----------- Email  ----------------------
{\small
{\tt{
\noindent
${}^a$umpei@phys.huji.ac.il\;\;\;\;\;
${}^b$maeda@waseda.jp
\\
}}
}

%%----------------------------------------
\end{center}

\vspace*{1.0cm}

%---------------------------------------------------------------------%
%---------------------------------------------------------------------%

%---------------------------------------------------------------------%

\begin{abstract}
Analyzing a capillary minimizing problem for a higher-dimensional extended fluid, we find that there exist startling similarities between the black hole-black string system (the Gregory-Laflamme instability) and the liquid drop-liquid bridge system (the Rayleigh-Plateau instability), which were first suggested by a perturbative approach. In the extended fluid system, we confirm the existence of the critical dimension above which the non-uniform bridge (NUB, i.e., {\it Delaunay unduloid}) serves as the global minimizer of surface area. We also find a variety of phase structures (one or two cusps in the volume-area phase diagram) near the critical dimension. Applying a catastrophe theory, we predict that in the 9 dimensional (9D) space and below, we have the first order transition from a uniform bridge (UB) to a spherical drop (SD), while in the 10D space and above, we expect the transition such that UB $\to$ NUB $\to$ SD. This gives an important indication for a transition in the black hole-black string system.
\end{abstract}

\end{titlepage}

%\preprint{arXiv:0803.3037 [hep-th], WU-AP/279/08}
%\title{ Liquid Bridges and Black Strings in Higher Dimensions }
%\author{ Umpei Miyamoto }
%\affiliation{ Racah Institute of Physics, Hebrew University, Givat Ram, Jerusalem 91904, Israel }
%\email{umpei@phys.huji.ac.il}
%\author{ Kei-ichi Maeda }
%\affiliation{ Department of Physics, Waseda University, Okubo 3-4-1, Tokyo 169-8555, Japan }

%\pacs{02.30.Xx, 03.65.Vf, 04.50.+h, 04.70.-s, 04.70.Bw, 68.35.Md, 11.25.-w}

% 02.30.Xx Calculus of variations
% 03.65.Vf Phases: geometric; dynamic or topological
%04.50.+h Gravity in more than four dimensions, Kaluza-Klein theory, unified 
%field theories, alternative theories of gravity (see also 11.25.M 
%Compactification and four-dimensional models), dilaton gravity 
% 05.70.Np Interface and surface thermodynamics (see also 68.35.Md Surface 
%thermodynamics, surface energies in surfaces and interfaces) 
%04.70.-s Physics of black holes(see also 97.60.Lf-in astronomy) 
%04.70.Bw Classical black holes  
%11.25.-w Strings and branes

%\maketitle

The most fundamental theory of nature, which is
 now believed to be string/M theory, predicts 
that spacetime dimension is higher than four. 
In four dimensional (4D) spacetime, we have a uniqueness theorem of
 a stable black hole (BH).
In higher dimensions, however, in addition to a BH,
we find a new phase of black objects such as a black string (BS) 
 and expect a transition between those 
objects. Its property may depend on the dimension. 
Hence, in order to understand the physics in higher dimensions, 
it is indispensable to reveal the mechanism 
of the transitions between the black objects.

In a spacetime in which one spatial dimension is 
compactified on a circle, for example,
there are three possible black ``objects", i.e.,
a localized black hole (LBH)~\cite{Harmark,KW}, 
a uniform black string (UBS), and a non-uniform black string (NUBS)
\cite{Gubser,KW,Sorkin1,Sorkin2}. Although the transitions 
between them may occur, 
our knowledge about those dynamics has been so far very limited.
The UBSs are unstable if the size of the compact dimension
is larger than some
critical length scale, as shown by Gregory and Laflamme (GL)~\cite{GL}. 
Its fate is, however, still controversial~\cite{dynamics}.
In 5D and 6D spacetimes, ``complete'' phase diagrams are available
\cite{KW,Sorkin2} (see~\cite{reviews} for reviews). 
From these phase diagrams, we
find that the UBS transits suddenly to the LBH as 
the compactification scale  gets larger than a critical size
\cite{Gubser,Kol2002}. On the other hand, 
for the case of the large spacetime dimension ($D\geq 14$),
we have another expectation
such that the phase transition occurs smoothly, i.e.,
the UBS is deformed to the NUBS continuously
~\cite{Sorkin1}.
Thus, our current interest is to figure out 
the complete phase structure and 
to study whether there exists a stable NUBS 
and how the unstable UBS
reaches NUBS.
%%%%%%%%%%%%%%%%%%%%%%%%%%%%%%%

To understand the dynamical
features of such ``black objects", 
the membrane paradigm of BHs~\cite{paradigm} as well as the BH 
thermodynamics~\cite{thermo} have been sometimes discussed.
Among them, there was a remarkable discovery by 
Cardoso, Dias and Gualtieri~\cite{RP1} that 
the GL instability and the Rayleigh-Plateau (RP) instability, 
which had been found in extended 
fluids~\cite{Plateau}, share many common features such 
as the dimensional dependence of the critical wavenumber
and that of the growth rate. 
They also pointed out that 
there may exist a critical dimension~\cite{RP1} similar to that of 
the BH-BS system~\cite{Sorkin1}.
%%%%%%%%%%%%%%%%%%%%%%%%%%%%%%%

For the stability and dynamics 
of the extended fluid system,
there have been extensive studies in the case of the usual 3D space. 
The equilibria of the liquid bridge/drop system
are obtained by the
variation of the surface area while keeping the volume fixed, 
which is called the \textit{capillary minimizing problem} (or \textit{Plateau's problem})~\cite{Fin86}. 
The stability and dynamics can be investigated by the surface diffusion 
equation~\cite{Mul57}, which does not increase the surface area~\cite{CT94}.
The spherical drop (SD) is stable~\cite{NM,EMS98}, while
the non-uniform bridge (NUB), which is called the
\textit{Delaunay unduloid}~\cite{unduloid}, is always
unstable~\cite{Vog,BBW98}. The uniform bridge (UB) becomes unstable
when the linear dimension is longer than its circumference~\cite{NM} 
(the RP instability). The dynamical collapse of
a long liquid bridge results in a (self-similar) drop formation~\cite{Eggers93,CFM}.
%%%%%%%%%%%%%%%%%%%%%%%%%%%%%%%

Here we generalize the capillary minimizing problem to 
the higher-dimensional spaces in which one dimension is bounded.
It is a mimic of higher dimensional space with one compactified dimension.
By extremalizing the surface area while keeping the volume fixed,
we obtain axisymmetric liquid equilibria non-perturbatively. 
Then, we construct the phase diagrams and compare 
the results with those of the BH-BS system.
 We find that the system of SD-(N)UB and the BH-BS system 
exhibit remarkable similarities.
Note that we should minimize
the liquid surface area, 
while maximize the area of event horizons, which
is proportional to the BH entropy. 
Climbing up the ``ladder" of dimensions, we find that
there exists a critical dimension 
above which the NUB can be a global minimizer of the area. 
In addition, we obtain a very interesting non-trivial structure of phase 
diagram near the critical dimension, 
which may tell us how the phase transition appears.
%%%%%%%%%%%%%%%%%%%%%%%%%%%%%%%

We consider an axisymmetric hypersurface (liquid surface) in 
$(n+2)$-dimensional flat space ($n\geq 1$),
in which one dimension in
the direction of the symmetric axis is bounded as
$ z \in [-L/2,L/2] $, where $z$ is the axis coordinate.
The surface is given by  $r=r(z)$,
where $r$ represents the radius of the $n$-sphere at $z$ of 
 the hypersurface. Then, 
the volume and surface area are given by the functionals of
$r(z)$ as
\begin{eqnarray}
&&
	V[r]
	=
	\Omega_{n} \int_{-L/2}^{L/2} dz\;r^{n+1}(z)
	\ ,
\nonumber
\\
&&
	A[r]
	=
	(n+1)\Omega_{n}
	\int_{-L/2}^{L/2}
	dz\;
	\sqrt{ 1 + r^{\prime 2}(z) }\; r^{n}(z) \ ,
\label{eq:area}
\end{eqnarray}
where $ \Omega_n = \pi^{(n+1)/2}/\Gamma( (n+1)/2 +1 ) $ is the volume of 
a unit $n$-sphere, and a prime denotes the $z$ derivative.
The equilibrium configuration will be obtained by extremalizing the action
$ I[r] := A[r] - \kappa V[r] $, where $\kappa$ is a Lagrange multiplier (which is geometrically the \textit{constant mean curvature} of the hypersurface). Thus, 
we have the following Euler-Lagrange equation,
\begin{eqnarray}
	r^{ \prime\prime }
	- n \left( 1 + r^{ \prime 2 } \right)/r
	+  \kappa \left( 1 + r^{ \prime 2 } \right)^{3/2}
	=
	0\ ,
	\label{eq:EL}
\end{eqnarray}
where the multiplier $\kappa$ is determined by 
keeping the volume constant.
This equation has two trivial analytic solutions; 
UB, i.e., the uniform bridge with the radius $r=r_0$ and
$\kappa=n/r_0$,
and SD of $r^2+z^2=R_0^2$, i.e. $(n+1)$-D sphere
with the radius $R_0$ and $\kappa=(n+1)/R_0$. 
\\ \indent
The static perturbations around the UB
gives a simple indication for the similarities 
between the present system and the BH-BS system
as shown in \cite{RP1}.
By setting 
$ r = r_0 + \epsilon r_1(z) $ and 
$ \kappa = n/r_0 + \epsilon\kappa_1 $ with $ \epsilon \ll 1 $,
and solving the linear-order perturbation equation 
$
	r_1^{\prime\prime}
	+ n r_1 / r_0^2  
	=
	- \kappa_1 \,
$
with the boundary conditions of $r_1^{\prime}|_{z=0}=0$  and  
$r_1^{\prime}|_{ \pm L/2 }=0$,
 we find 
$
	r_{1}
	=
	r_{1}^{(0)} \cos( \sqrt{n} \, z / r_0 ) \ ,
$
where $r_{1}^{(0)}$ is an integration constant, and $ \kappa_1 =0$
from the volume constraint.
This static solution 
gives the critical mode of the RP instability.
We find the critical radius of UB as
$r_{\rm cr} =\sqrt{n} L/2\pi$ below which
the UB is unstable.
This sinusoidal onset mode and its dimension 
dependence~\cite{RP1} are quite similar to those of the GL 
instability~\cite{Kol:2004pn}.
%%%%%%%%%%%%%%%%%%%%%%%%%%%%%%%

Now we show the existence of a critical dimension beyond which the NUBs 
become stable.
The present system with (\ref{eq:area}) has a scale invariance,
that is, we find one parameter family of solutions by rescaling 
such as
$L\rightarrow \beta L$,
$r\rightarrow \beta r$,
$z\rightarrow \beta z$.
So we introduce a reduced 
volume and an area by
\begin{eqnarray}
	\mu
	:=
	V/L^{n+2} \ ,
	\;\;\;\;\;
	a
	:=
	A/V^{ (n+1)/(n+2) }\ ,
\end{eqnarray}
which are invariant under the rescaling of length.
We denote the quantities associated with the 
critical uniform bridge by the subscript cr as $\mu_{\rm cr}$.
Note that $a$  for a spherical drop, $a_{\mathrm{SD}}$, keeps
constant as $\mu$ changes.
We have numerically obtained one parameter family
of NUB solutions,
using the first integral of Eq.~(\ref{eq:EL}), i.e.
\begin{eqnarray}
	{1\over 2}r^{\prime 2}+U(r)
	=
	0\ ,
\;\;
	U(r)
	:=
	{1\over 2}
	\left[ 1-{(n+1)^2 r^{2n} \over \kappa^2
        ( r^{n+1} + C )^2 } \right]\ ,
\label{eq:EL2}
\end{eqnarray}
where $C$ is an integration constant,
and introducing the parameter
$\alpha:=r_-/r_+ $ which is the ratio of the smallest radius ($r_-$)
of NUB to the maximum one ($r_+$).
A series of $\alpha$ may describe a transition 
from the UB to the SB because $\alpha=1$ and 0 correspond to
the UB and SD, respectively.
%----------------------------------------------------------------------%
\begin{center}
	\begin{figure*}[t]
		\setlength{\tabcolsep}{ 12 pt }
		\begin{tabular}{ c  c c c  c }
			$ \mu/ \mu_{\rm cr}  = 1.00$ &
			$ \mu/ \mu_{\rm cr}  = 0.980$ &
			$ \mu/ \mu_{\rm cr}  = 0.676 $ &
			$ \mu/ \mu_{\rm cr}  = 0.309 $ &
			$ \mu/ \mu_{\rm cr}  = 0.279 $ 
			\\
			\includegraphics[width=2.3cm]{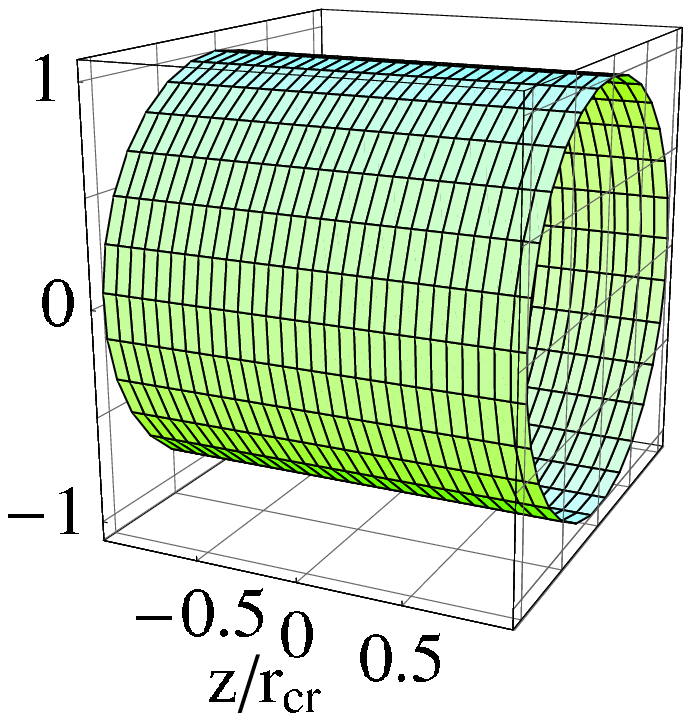} &
			\includegraphics[width=2.3cm]{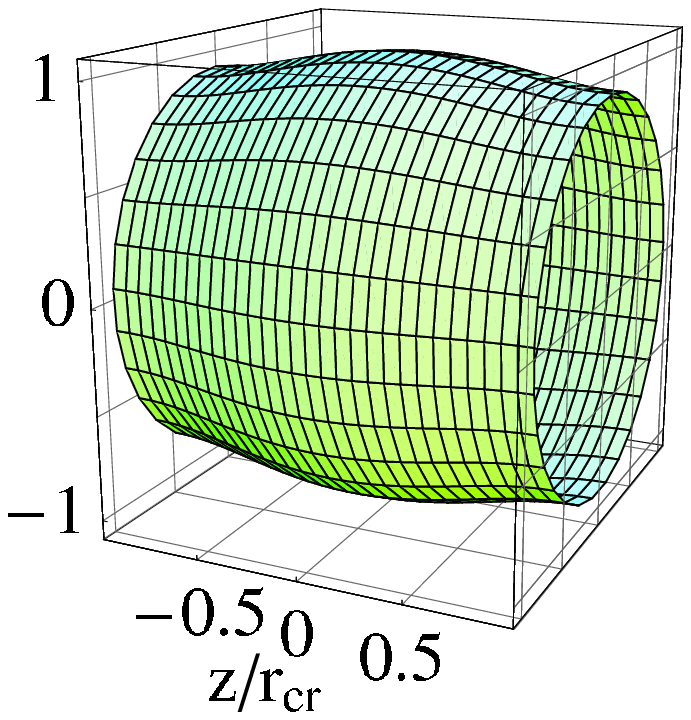} &
			\includegraphics[width=2.3cm]{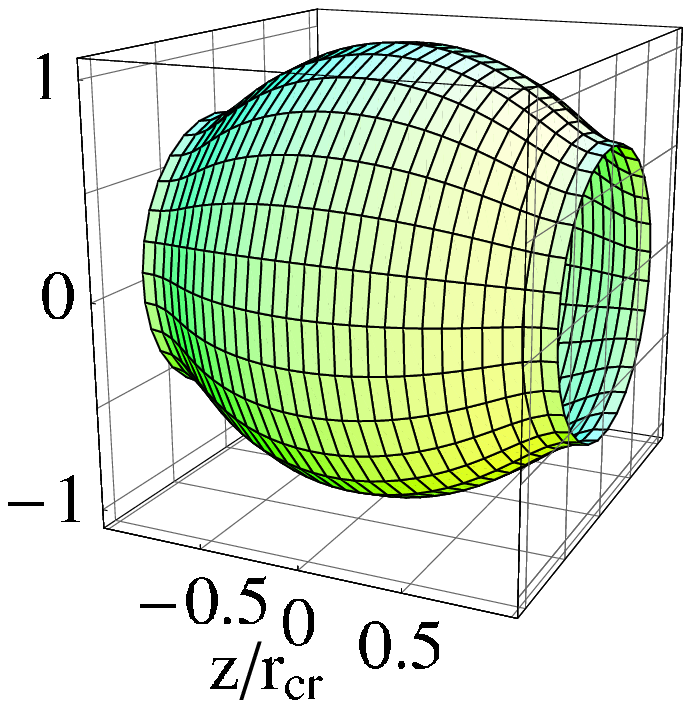} &
			\includegraphics[width=2.3cm]{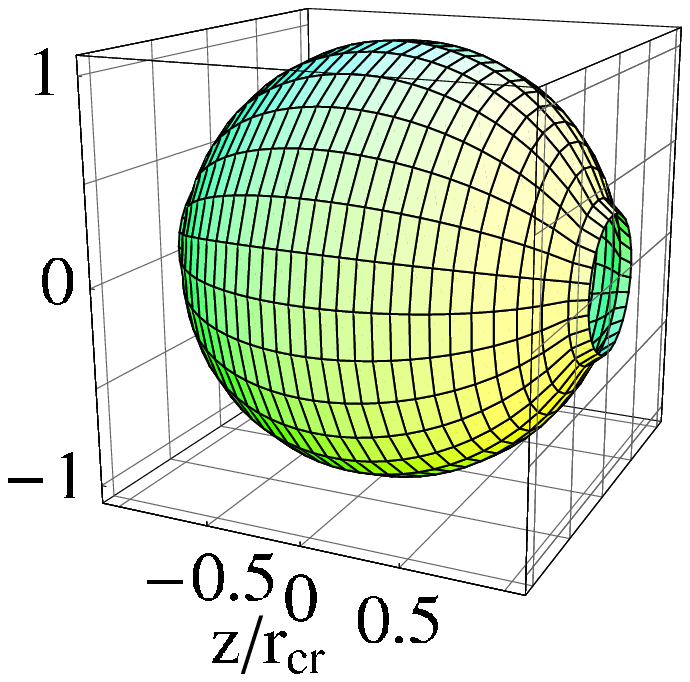} &
			\includegraphics[width=2.3cm]{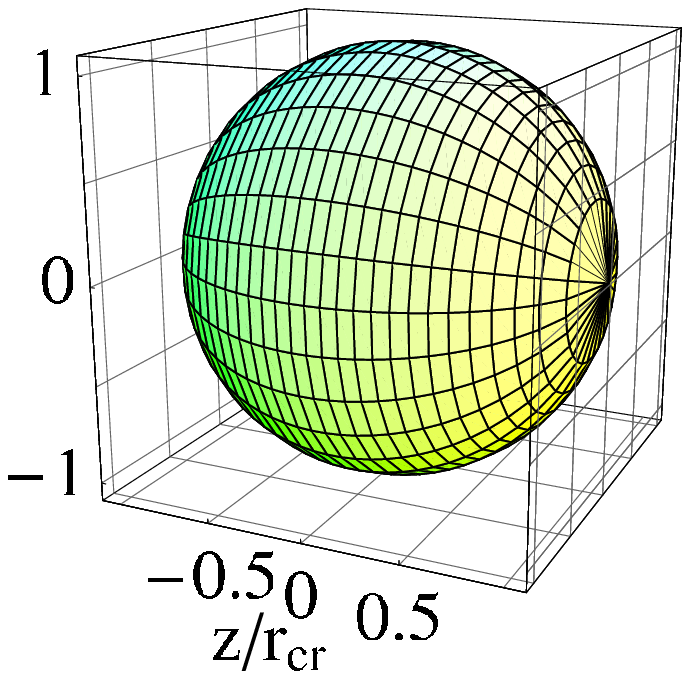} 
		\end{tabular}
	\caption{
	\footnotesize{
A uniform bridge (UB), non-uniform bridges (NUBs), and
a spherical drop (SD) at the transition in 12D.
$\mu$ and $\mu_{\rm cr}$ are
the reduced volume defined in the text and its critical value. 
$r_{\rm cr} $ is the radius of the critical uniform bridge.
The transition occurs from the left to the right as
$L$ increases while keeping the volume fixed.
The SD appears at $ \mu/ \mu_{\rm cr} = 0.279 $
$(< \mu_{\rm touch}/ \mu_{\rm cr}=0.330)$ via the first order transition.}
			}
			\label{fg:config}
	\end{figure*}
\end{center}
%----------------------------------------------------------------------%
First, we show the results for two typical dimensions;
 4D and 12D, which are
representatives for lower dimensions ($1\leq n \leq 6$) and 
higher ones ($n \geq 10$), respectively. The transition range of dimensions
is $7\leq n \leq 9$, which we will discuss later. 
Several equilibrium configurations for 12D
 embedded in a schematic 3D space are shown 
in Fig.~\ref{fg:config}. 
We also show the volume-area diagrams for 4D and 12D
in Fig.~\ref{fg:ma}.
In both dimensions, as $\alpha$ decreases to zero,
the NUB reaches 
the SD which ``touches itself'' at the boundary $ z = \pm L/2 $
when $ \mu = \mu_{ \mathrm{ touch } } $. 
For the 4D space, we find that
$\mu$ increases as the NUB smoothly approaches the critical drop,
and the NUBs exist only in the larger area region compared with
those of SDs and of UBs.
Hence we suspect that these NUBs are unstable like those in 3D case
~\cite{Vog,BBW98}.
This will be justified when we adopt catastrophe theory (see later).
For the 12D case, however, the NUB appears
in the region with smaller area than those of SDs and of UBs,
and it approaches a critical drop with a small 
cusp structure (see Fig.~\ref{fg:ma}).
We expect these NUBs (except for the upper branch of a cusp structure)
are stable because the areas are smallest for a given volume 
$V$ and length $L$.
 This is also confirmed by the argument of catastrophe.
This difference depends on whether the volume-area curves
of SDs and UBs intersects (4D) or not (12D).
%----------------------------------------------------------------------%
\begin{center}
	\begin{figure}[b]
		\setlength{\tabcolsep}{ 28 pt }
		\begin{tabular}{ cc }
			\includegraphics[width=6cm]{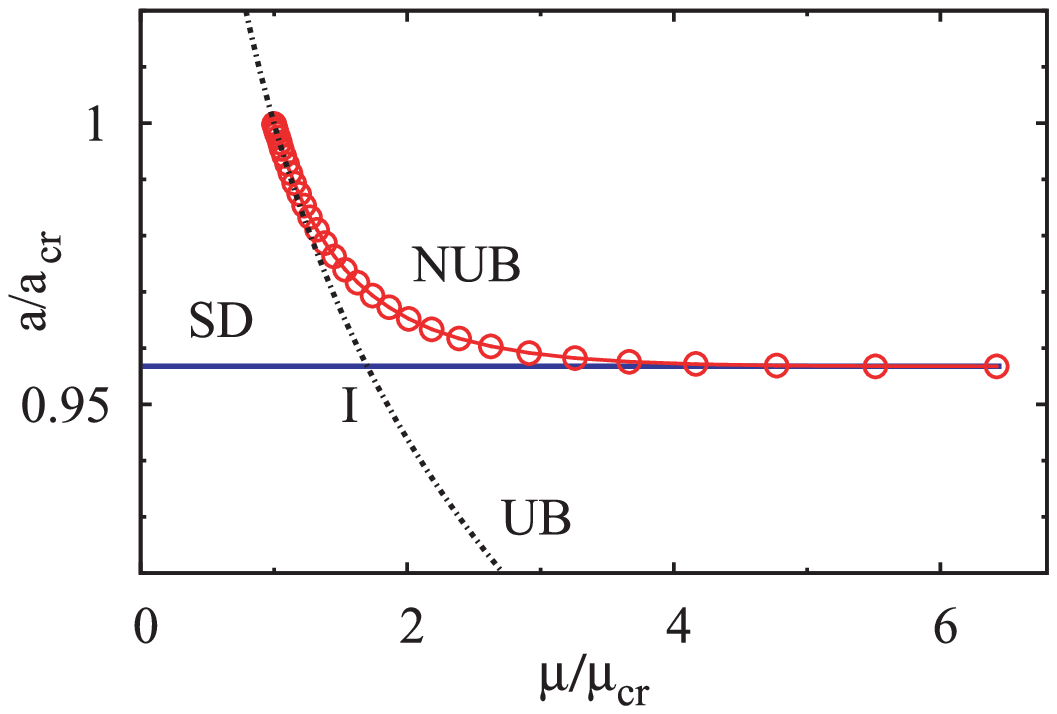} &
			\includegraphics[width=6cm]{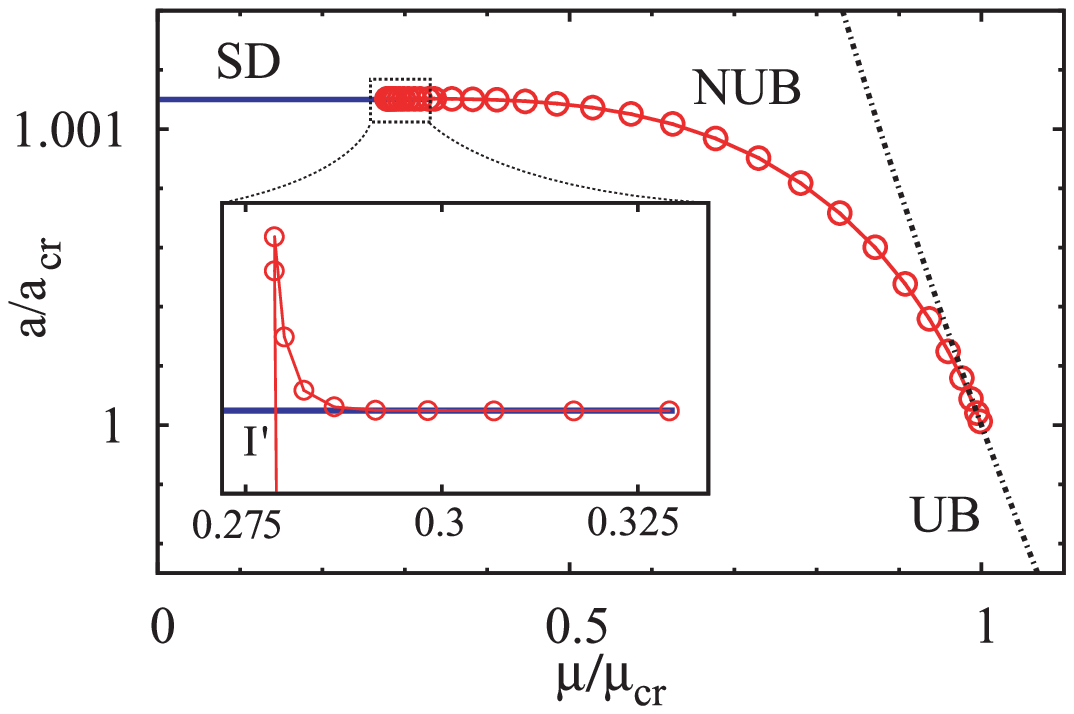}
		\end{tabular}
		\caption{
		\footnotesize{
The volume-area diagrams for 4D (left) and 12D (right), including the uniform bridge (UB), non-uniform bridge (NUB) and spherical drop (SD). 
We enlarge the diagram near the SD in 12D,
where we find a cusp structure.
The solution with the smallest area for a given volume is energetically favored.
Hence the UB in 4D and the NUB in 12D transit to the SD at the intersecting
points, I and I$'$, respectively. }
		}
		\label{fg:ma}
	\end{figure}
\end{center}
%----------------------------------------------------------------------%
%----------------------------------------------------------------------%
\begin{center}
	\begin{figure}[h]
		\setlength{\tabcolsep}{ 28 pt }
		\begin{tabular}{ cc }
%			\hspace{-.3cm}
			\includegraphics[width=6cm]{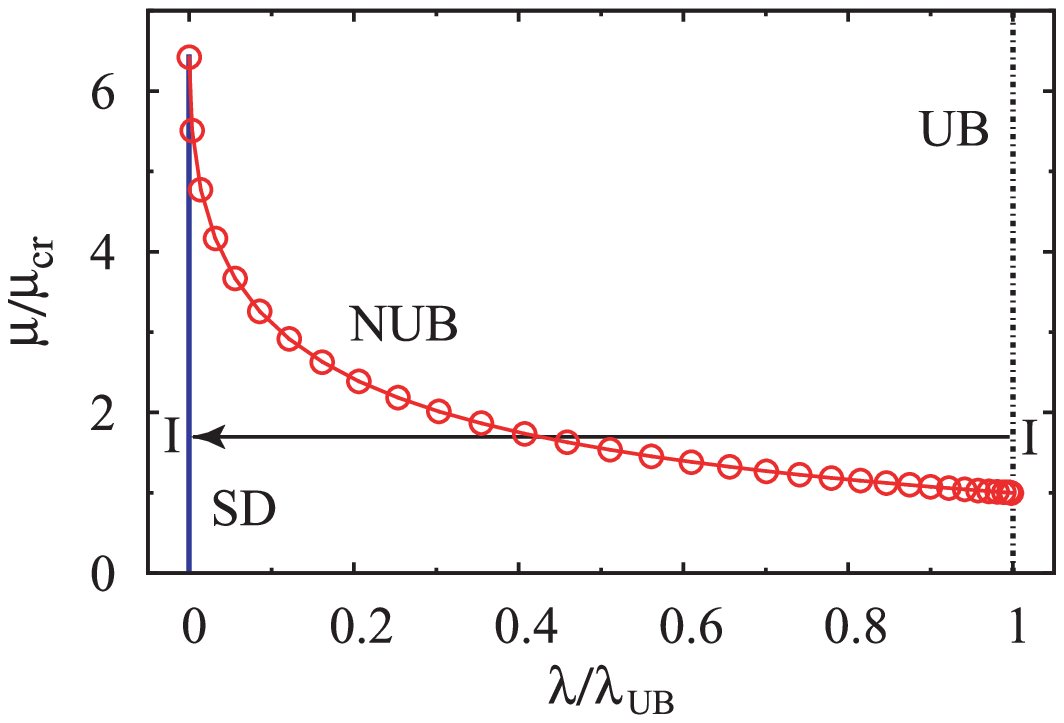} &
			\includegraphics[width=6cm]{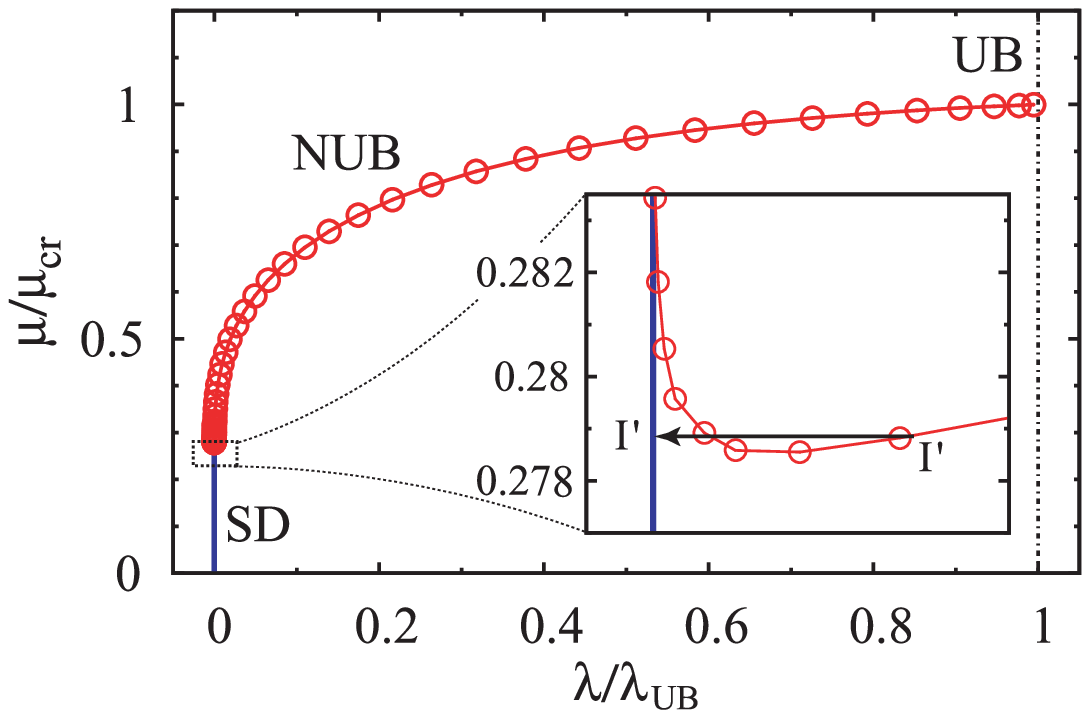}
		\end{tabular}
		\caption{
		\footnotesize{
The tension-volume diagrams for 4D (left) and 12D (right).
In 4D, the UB transits to the SD with a finite jump of $\lambda$
as the arrow points, while in 12D, the transition from the UB to 
the NUB is continuous. The transition to the SD is, however,
discontinuous although the jump is very small (see Fig.~\ref{fg:ma}).}
		}
		\label{fg:lm}
	\end{figure}
\end{center}
%----------------------------------------------------------------------%

Adopting the least area law,
we can predict the most favored (and probably stable)
state from the volume-area diagrams in Fig.~\ref{fg:ma}.
Consider the case fixing the volume and stretching the length $L$,
i.e., decreasing $\mu$. 
In 4D, because the area of NUB is always larger than others,
as $\mu$ decreases, 
the favored state shifts from the UB to the SD at their intersecting 
point I with a discrete transition of the  configuration. 
This phase diagram is qualitatively 
similar to the mass-entropy diagram of the 6D BH-BS system 
(Fig.~6 in Ref.~\cite{KW}). 
Note that there are a few differences, that is, 
one is the physical role of the 
area mentioned before, 
and the other is that the SDs do not deform unlike 
the LBHs, which stems from the neglect of gravity. 
While in 12D, the
favored state shifts from the UB to the NUB successively, and 
eventually transit to the SD. 
This transition to the NUB is of second order or higher. 
However, there exists a small cusp near the SD.
So the transition to the SD occurs at the point
I$'$, where the two curves of NUB and of SD
intersect. This transition is of first order.
We expect that there also exists this type of cusp
in the higher dimensions, although we could only confirm it
numerically until 15D because of difficulty
of numerical accuracy.

Here we introduce an order parameter of the present system
in order to compare our phase structure with that  
in the gravity side. We find that the following parameter,
\begin{eqnarray}
	\lambda
	:=
	\frac{ L }{ A }
	\left( \frac{ \partial A }{ \partial L } \right)_V
        =-(n+2){\mu\over a}{\partial a\over \partial \mu}
\,,
\end{eqnarray}
which we call the \textit{reduced tension},
can be an appropriate order parameter.
This $\lambda$ is analogous to the relative tension in the 
BH-BS system~\cite{tension}.
$\lambda$ is constant for UB as $\lambda_{\rm UB}=1/(n+1)$, 
 and it vanishes for SDs. 
For NUBs, $\lambda$ is evaluated
from the gradient of the $\mu$-$a$ curve.
%----------------------------------------------------------------------%
In Fig.~\ref{fg:lm}, we show $(\lambda,\mu)$ diagrams.
We again find  a qualitative agreement between the $(\lambda,\mu)$ diagram
for the 4D space and 
the tension-mass diagram of BH-BS system in the 6D spacetime (see Fig.~5 in 
Ref.~\cite{KW}), which had been originally conjectured in~\cite{Kol2002}. 
As the control parameter $\mu$ decreases, the UB
transits to the SD at
the intersecting point,
$ \mu = \mu_{\mathrm{ I }} $ $ ( > \mu_{\rm cr} ) $,
with a finite jump $ \Delta \lambda = 1/3 $ $ ( = \lambda_{\mathrm{UB}} ) $. 
While in 12D, the UB transits to the NUB continuously.
The transition from the NUB to the SD at the point I$'$ is discontinuous 
with a very small jump
$\Delta \lambda \simeq 1.1 \times 10^{-7} $ ($ \ll \lambda_{\mathrm{UB}}$).
%$\Delta \lambda / \lambda_{\mathrm{UB}} \ll 1$.
%----------------------------------------------------------------------%
%----------------------------------------------------------------------%
\begin{table}[t]
	\begin{center}
	\caption{
	\footnotesize{
The expected transition patterns in terms of space dimensions.
The first (``second") order $\&$ first order means that
 we have the first  (``second") order
transition for UB $\rightarrow$ NUB, and then 
the first order transition for NUB $\rightarrow$ SD.
The ``second" means that it may be of second order but might be 
the higher order transition. For much higher dimensions than 15, 
we have not confirmed the existence of a cusp because of numerical
difficulty, but we may expect that it 
is of first order because the topology change
is associated in this transition. }
}
\label{table}
\vspace{.3cm}
\footnotesize{
\setlength{\tabcolsep}{3pt}
\begin{tabular}{ | l|| c | c | c | c  | c  |}
	\hline \hline
	space dimension ($n+2$)	&	3 ~$\sim$~ 8	&	9	&	~~~~10~~~~ &	11	&	12	$\sim$
	\\
\hline \hline
	transition type	&
	\multicolumn{2}{|c|}
	{~~UB\hskip .5cm $\longrightarrow$ \hskip .5cm	SD ~~}	&
	\multicolumn{3}{|c|}
	{UB\hskip .5cm $\longrightarrow$\hskip .5cm  NUB  
	$\hskip .5cm \longrightarrow$ \hskip .5cm  SD}
	\\
\hline
	order of transitions &\multicolumn{2}{|c|}{first order} &
	\multicolumn{2}{|c|}
	{first order \& first order}	&
	``second"	order \& first order
	\\
\hline
	number of cusps in $\mu$-$a$ diagram &~~no cusp~~ &
	\multicolumn{3}{|c|}
	{two cusps } &
	one cusp \\
\hline\hline
	\end{tabular}
}
	\end{center}
\end{table}
%----------------------------------------------------------------------%
%----------------------------------------------------------------------%

%----------------------------------------------------------------------%
\begin{center}
	\begin{figure}[h]
		\setlength{\tabcolsep}{ 28 pt }
		\begin{tabular}{ cc }
			\hspace{-.25cm}
			\includegraphics[width=6cm]{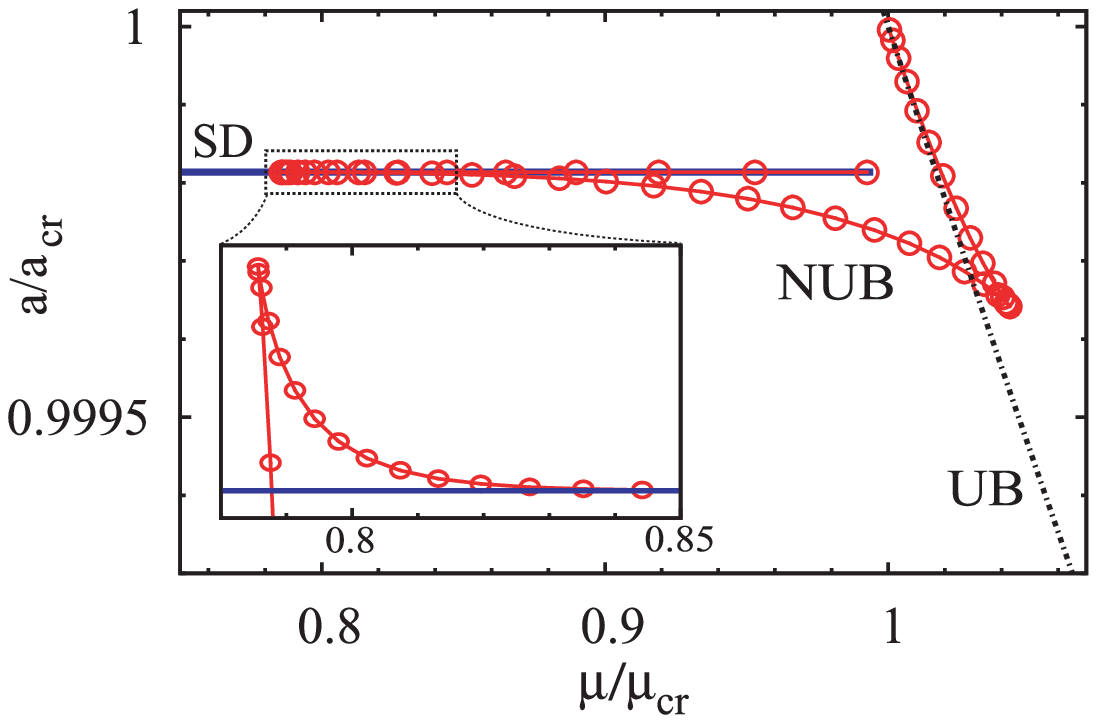} &
			\includegraphics[width=4.5cm]{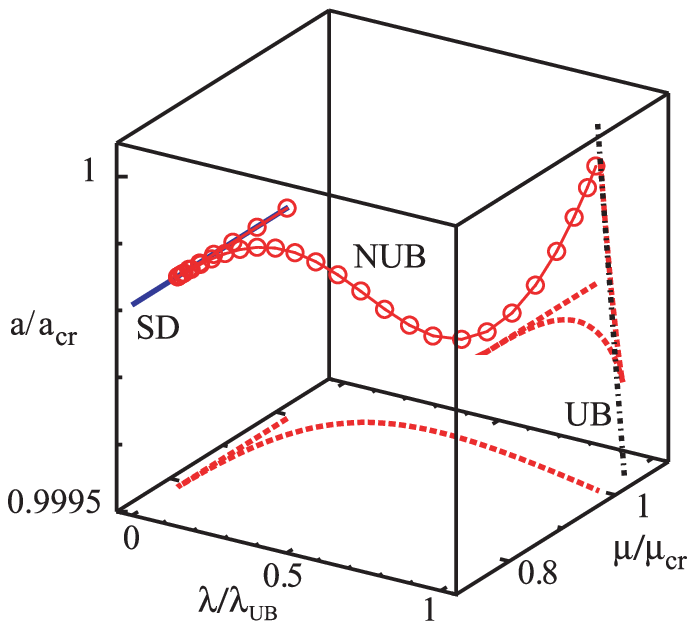} 
		\end{tabular}
			\caption{
			\footnotesize{
The volume-area and tension-volume-area diagrams for 10D. 
Although we find two cusp structures in the ($\mu, a$) diagram, 
the curve of the NUB in the 3D phase space is smooth, and connects the UB and the SD
at its both ends.}
			}
			\label{fg:10D}
	\end{figure}

\end{center}
%----------------------------------------------------------------------%
Now, we discuss the behaviors in the transition
range of dimensions (9D, 10D, and 11D spaces).
In Fig.~\ref{fg:10D}, we show the $(\mu,a)$ diagram
and the three dimensional $(\lambda,\mu,a)$ diagram for 10D space.
We find that the NUB has two cusp structures in the $(\mu,a)$ plane.
The cusp near the SD is the same as that found in 12D.
A new cusp also appears near the UB.
The upper branch (the larger area)
corresponds to the curve of the NUB which appeared in 4D, while
the lower branch (the smaller area) corresponds
to the curve of the NUB found in 12D.
Because we find the cusp structure, 
we can adopt the catastrophe theory
by setting $a$, $\mu$, and $\lambda$ to be
a potential function, a control parameter, and
a state variable, respectively~\cite{catastrophe}.
We expect that the lower branch is stable 
while the upper one is unstable.
This expectation is justified from the fact that 
the area in the upper branch is always larger than that of the UB,
which means that the NUBs are unstable against
small perturbations just as the 3D case.
Such interpretation is consistent with 
our expectation about stabilities in 4D and 12D.
The Landau-Ginzburg argument also can be applied~\cite{Kol2002}.
%%%%%%%%%%%%%%%%%%%%%%%%%%%%%%%
The similar cusp structures are observed in 9D and 11D spaces. 
In the 9D case, however, such a structure appears always in
the larger area region than that of UBs.
So this has nothing to do with a realistic transition.
%%%%%%%%%%%%%%%%%%%%%%%%%%%%%%%

We summarize the results in Table \ref{table}.
The results strongly support the 
existence of the critical dimension on the gravity side~\cite{Sorkin1} and 
fills a ``missing link'' in the phase diagrams of the BH-BS system,
unknown except 5D and 6D~\cite{Kol2002,KW}.

Finally, we should stress that all equilibria obtained here are static
solutions of the surface diffusion equation~\cite{Mul57}. 
Although we could analyze the stability from the second variation
of the area~\cite{Vog}, one can also examine the 
stability and the liquid bridge/drop transition process 
by solving the dynamical equation. What is the
most interesting is whether or not we
confirm the local/global convergence to the NUBs for  
$n\geq 8$.
Such analysis would be a footstep to understand the 
dynamics of the black strings in higher dimensions.

We would like to thank O.J.C.~Dias for suggesting a possible extension of 
his work, and B.~Kol and R.~Emparan for fruitful discussions. U.M.~is supported by the 
Golda Meir Fellowship, by The Israel Science Foundation grant No.607/05, and 
by the DIP grant H.52. This work is partially supported 
by the Grant-in-Aid for Scientific Research
Fund of the JSPS (No.19540308) and for the
Japan-U.K. Research Cooperative Program,
and by the Waseda University Grants for Special Research Projects and
 for the 21st-Century
COE Program at Waseda University.

\textit{Note added ---}
After the first submission of a manuscript to arXiv, one of the present authors wrote an article~\cite{M} in which the linear and non-linear dynamics of the RP instability were investigated in the framework of the surface diffusion. In particular, it was shown that the NUB indeed serves as an endpoint of the RP instability around the critical dimension and above.
%----------------------------------------------------------------------%
%----------------------------------------------------------------------%

%----------------------------------------------------------------------%
%----------------------------------------------------------------------%

\end{document}